\documentclass[lettersize,journal]{IEEEtran}
\usepackage{amsmath,amsfonts}
\usepackage{algorithmic}
\usepackage{algorithm}
\usepackage{array}
\usepackage[caption=false,font=normalsize,labelfont=sf,textfont=sf]{subfig}
\usepackage{textcomp}
\usepackage{stfloats}
\usepackage{url}
\usepackage{verbatim}
\usepackage{graphicx}
\usepackage{threeparttable}
\usepackage{booktabs}
\usepackage{multirow}
\usepackage[inkscapelatex=false]{svg}
\usepackage{enumitem}

\def\endthebibliography{%
  \def\@noitemerr{\@latex@warning{Empty `thebibliography' environment}}%
  \endlist
}
\makeatother

\usepackage{pifont}

\newcommand{\prjname}{ETHEREAL}

\begin{document}

\title{\vspace*{-2mm}\prjname: A 25.6-$\mu$s/inf. Low-latency Event-driven Graph-neural-network Processor \\ for High-resolution Vision at the Edge}

\author{Adrian Kneip,~\IEEEmembership{Member, IEEE}, Martin Lefebvre,~\IEEEmembership{Member, IEEE}, Daniel Gehrig,~\IEEEmembership{Member, IEEE}, \\ Victoria Catalán Pastor,~\IEEEmembership{Graduate Student Member, IEEE}, Davide Scaramuzza,~\IEEEmembership{Senior Member, IEEE}, \\ Marian Verhelst$^*$,~\IEEEmembership{Fellow, IEEE} and Charlotte Frenkel$^*$,~\IEEEmembership{Member, IEEE}
\vspace*{-4mm}

\thanks{Manuscript received March XX, 2026.}
\thanks{Authors with * share equal contribution. A. Kneip is with the Microelectronics Department (EEMCS Faculty), Delft University of Technology, 2628 CD Delft, Netherlands, and with the Department of Electrical Engineering, KU Leuven, 3000 Leuven, Belgium.
M. Lefebvre and C. Frenkel are with the Microelectronics Department (EEMCS Faculty), Delft University of Technology, 2628 CD Delft, Netherlands.
M. Verhelst is with the Department of Electrical Engineering, KU Leuven, 3000 Leuven, Belgium.
V. Catalán Pastor and D. Scaramuzza are with the Robotics and Perception Group, University of Zurich, Zurich, Switzerland.
D. Gehrig is with the University of Pennsylvania, PA 19104, Philadelphia, United States of America.
(Corresponding author: a.kneip@tudelft.nl).}
}


\markboth{}
{Kneip \textit{et al.}: ETHEREAL: A 25$\mu$s/inf. event-driven GNN Processor Chip}


\maketitle

\begin{abstract}


Dynamic vision sensors (DVS) are enticing candidates to reach the low-latency, sub-ms target of edge-vision applications, as they generate events with a $\mu$s-level time resolution. However, using DVS front ends also calls for novel algorithm/hardware back ends capable of efficiently handling  streams of sparse spatiotemporal events. While event-driven graph neural networks (EV-GNNs) have emerged as a solution on the algorithmic side that is both accurate and efficient, there is no dedicated hardware to date capable of efficiently supporting their mixed requirements of dense-regular compute operations and sparse-irregular memory accesses.
We therefore introduce ETHEREAL, the first EV-GNN processor chip, capable of bridging this gap by means of a neighbor-parallel spline-convolution engine combined with a split-2D/3D memory hierarchy that introduces a novel spatiotemporal event-caching mechanism. 
Measurement results demonstrate a 25.6$\mu$s latency and a 1.6$\mu$J energy per end-to-end event-wise inference on the state-of-the-art DAGr-GNN workload and VGA-resolution (640$\times$480 pixels) DSEC dataset.

\end{abstract}


\section{Introduction}
\label{sect:intro}

\begin{figure}[!t]
    \centering
    \includegraphics[width=\linewidth]{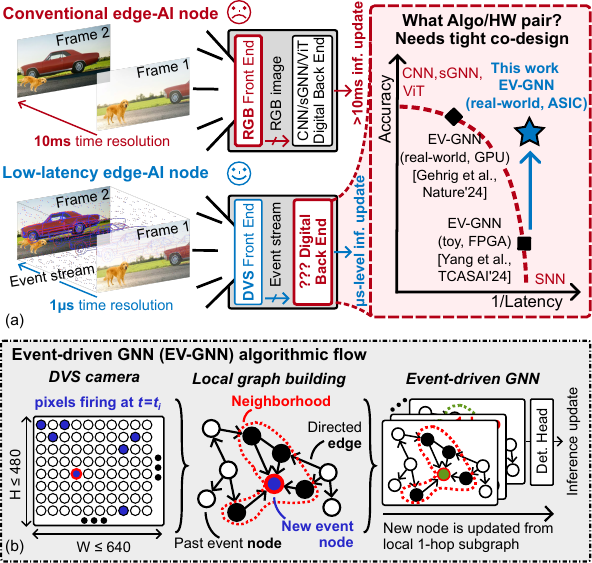}
    \caption{(a) Frame- versus event-based edge-vision platforms, and gap in the state of the art. (b) Working principle of EV-GNNs.}
    \label{Fig_intro}
    \vspace{-0.2cm}
\end{figure}

\IEEEPARstart{L}{ow} latency is a key ingredient toward the large-scale deployment of smart edge-vision systems with sub-ms real-time detection constraints, including safety-critical applications such as autonomous driving and drone navigation \cite{Maqueda2018,Gupta2021}, as well as virtual reality  \cite{Elbamby2018}.
Nowadays, typical edge-vision hardware (HW) nodes consist of a sensing front end followed by a processing back end, as illustrated in Fig.~\ref{Fig_intro}(a).
Commonly, these nodes capture scenes using standard RGB cameras, whose output frames are processed by a dedicated HW back end running a task-specific computer-vision algorithm such as a CNN for car detection \cite{cai2016unified}. As such, the latency of these systems is capped by the $\sim$10ms temporal resolution of standard frame-based RGB cameras \cite{camera_2026}.
In contrast, dynamic vision sensors (DVS) are capable of probing changes in a scene with a pixel-wise temporal resolution in the order of 1$\mu$s \cite{Lichtsteiner2008}. Yet, specific algorithms and processing back ends are required to efficiently leverage the fine-grained temporal resolution of the DVS event stream, combined with its high spatiotemporal sparsity.

On the algorithmic side, most existing event-based processing techniques face a trade-off between detection accuracy and latency per inference. On the one hand, event histograms \cite{cannici2019asynchronous, messikommer2020event}, time surfaces \cite{sironi2018hats}, vision transformers \cite{gehrig2023recurrent} or 3D \textit{synchronous} (or \textit{static}) graph neural networks (sGNNs) \cite{chen2024survey} allow processing data in a frame-like manner, thereby enabling batched execution on GPUs \cite{markidis2018nvidia} or dedicated accelerators \cite{chen2019eyeriss, dumoulin2024enabling, kneip2023impact, dong202528nm}. However, these standard methods use temporal aggregation, which negates the low-latency nature of the DVS front end. 
On the other hand, fully event-driven algorithms such as spiking neural networks (SNNs) can genuinely harness the low-latency nature of the event stream by running on dedicated hardware \cite{Viale2021,Frenkel2022}. However, as of today, their complex, time-explicit training process prevents them from achieving top accuracies on high-resolution detection tasks \cite{ottati2023_spiking}.

Recently, event-driven graph neural networks (EV-GNNs) have emerged as an enticing solution to bridge the accuracy-latency gap. At training time, EV-GNNs train on 3D-aggregated graph data to achieve high accuracy, similar to sGNNs. At inference time, however, EV-GNNs process the event stream \textit{asynchronously} to leverage local spatiotemporal update rules \cite{schaefer2022aegnn}, as illustrated in Fig.~\ref{Fig_intro}(b). First, they insert each new DVS event as a \textit{node} in the 3D spatiotemporal graph, connecting it through a set of \textit{directed edges} to its \textit{neighborhood} (NB) of past nodes through a distance-dependent graph-building process. Then, they process the local subgraph of this event with a GNN that relies on local message passing, thereby vastly reducing the number of operations compared to a synchronous pass on the entire graph. As such, EV-GNNs promise a high-accuracy, low-latency prediction update through event-driven computing.

Yet, no hardware to date could demonstrate these properties, as the few existing accelerators for EV-GNNs on FPGA \cite{Yang2025, Jeziorek2025, Liu2026} are currently limited to toy, low-resolution ($\leq$ 128$\times$128 pixels) setups. These designs cannot be scaled up to high-resolution ($\geq$ 640$\times$480 pixels) EV-GNN workloads \cite{Gehrig2024}, as they fail to concurrently leverage two aspects: on the one hand, the high spatiotemporal sparsity of the event stream, which results in irregular memory accesses; on the other hand, the locally dense-and-regular operations required for message passing, which calls for a high parallelism.

In this work, we address these concerns by introducing ETHEREAL, the first EV-GNN processor chip, capable of harnessing detection tasks with a high spatial resolution, such as the DSEC dataset for automotive detection (640$\times$480 DVS resolution) \cite{gehrig2021dsec}. Exploiting HW-SW co-design, ETHEREAL introduces HW support for new, precision-flexible computing kernels, as well as a novel memory hierarchy that is suited to the varying sparsity and data types of advanced EV-GNN architectures, such as DAGr-GNN \cite{Gehrig2024}. Measurement results on this workload showcase a 25.6$\mu$s latency and a 1.6$\mu$J energy per end-to-end event inference for DSEC, a \mbox{first-in-class} performance demonstration.

This paper is organized as follows. Section~II introduces the necessary background on state-of-the-art EV-GNNs and highlights their core challenges. Section~III introduces the ETHEREAL chip, which addresses these challenges. Sections~IV and V then respectively address datapath and memory innovations. Section VI finally presents and discusses measurement results, as well as perspectives.

\section{Background and Motivation}
\label{sect:background}
In contrast with mainstream AI algorithms (e.g., CNNs and transformers) that involve dense, tile-based operations, EV-GNNs rely on local data exchanges between sparsely connected neighbors to minimize the number of operations per prediction update \cite{schaefer2022aegnn, Yang2025, Gehrig2024}. Below, we identify the key properties of modern EV-GNNs, from a single layer to a full network, and translate them into HW challenges.

\begin{figure}[!t]
    \centering
    \includegraphics[width=\linewidth]{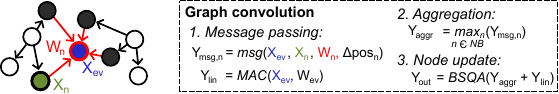}
    \caption{Illustration of graph convolution in modern EV-GNNs.}
    \label{Fig_graph_conv}
    \vspace{-0.2cm}
\end{figure}

 \begin{figure}[!t]
    \centering
    \includegraphics[width=\linewidth]{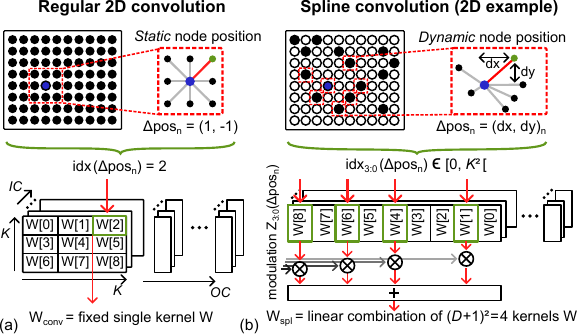}
    \caption{Comparison between weight kernels for (a) a regular 2D convolution on a pixel grid, interpreted as a dense graph, and (b) a spline convolution with a 2D-spatial spline basis. Illustrated for kernel size $K=3$ and spline degree $D=1$. $IC$ and $OC$ stand for input- and output-channel dimensions.}
    \label{Fig_SPL_conv}
    \vspace{-0.2cm}
\end{figure}

\subsection{Graph Convolution in Modern EV-GNNs}
Graph convolution is at the core of the EV-GNNs, enabling an efficient update of the graph by means of local message passing between the graph's sparsely interconnected nodes.

\vspace{0.2cm}

\subsubsection{Basics of Graph Convolution} A vanilla graph-convolution operation follows three steps, described in Fig.~\ref{Fig_graph_conv}. First, (1) a \textit{message-passing} step transforms the event- and neighbor-node features ($X_{ev}$ and $X_n$, respectively, where $n \in [0, NB[$) into messages $Y_{msg, 0:NB-1}$ by means of spline-modulated weights $W_{spl}$. Optionally, the (relative) distance between the event node and its neighbor, $\Delta pos_n = (dx, dy)_n$, or a transformation thereof, can be concatenated to the different input tensors to leverage spatial information across the neighborhood \cite{Yang2025}. Second, (2) an \textit{aggregation} step combines these messages and extracts the maximum value $Y_{aggr}$ per feature channel. With directed edges, information flows in a causal manner from the neighbors towards the new node during aggregation, thereby preventing the costly spread of node modifications seen in bidirectional EV-GNNs \cite{schaefer2022aegnn}. Third, (3) a \textit{node-update} step first adds a \textit{self-linear} transformation $Y_{lin}$ of the event-node's features $X_{ev}$ to the aggregated result, before applying bias, scale, quantization and activation (BSQA) to obtain the graph-convolution output $Y_{out}$ of the event node at layer $L$. This output is not only sent to the next layer, but also updates the graph-data map stored in the system's memory.

\vspace{0.2cm}

\subsubsection{Spline Convolution} To turn positional information into significant accuracy benefits, more complex message-passing operations have been proposed. 
Among them, spline convolution \cite{Matthias2018} can be interpreted as a variant of regular 2D convolution suited for sparse, irregularly distributed data, as depicted in Fig.~\ref{Fig_SPL_conv}. During regular 2D convolution, a single weight is statically applied to each neighbor location within a kernel window of size $K \times K$, yielding a message
\begin{equation}
    Y_{msg, n, j} = \sum_{i=0}^{IC-1}{X_{n,i} \, W_{conv, i, j}[n]},  \, n\in [0, K^2[ \,,
\end{equation}

\noindent where $i \in [0, IC[$ and $j \in [0, OC[$ respectively stand for the input and output channel. This operation assumes a regular and predetermined distance between a node and its neighboring pixel $n$ (Fig.~\ref{Fig_SPL_conv}(a)), which does not match with the irregular neighbor-to-event distance in graphs, nor with their sparsity. 
In contrast, spline convolution accounts for dynamic position changes by applying different spline weights $W_{spl}$ based on the neighbor-to-event distance $\Delta{pos_n}$, giving

\begin{equation}
    Y_{msg, n, j} = \sum_{i=0}^{IC-1}{X_{n,i} \, W_{spl, i, j }(\Delta{pos_n)}, \, n\in[0, NB[ \,,}
    \label{Eq_spl_conv_1}
\end{equation}

\noindent where the spline weight $W_{spl}$ is obtained as the linear combination of $(D +1)^2 = 4$ learnable weights $W$ (Fig \ref{Fig_SPL_conv}(b)), where $D=1$ is the spline's \textit{degree} \cite{Matthias2018}. These four $W$ tensors are selected out of a kernel of $K\times K = 9$ tensors based on the $\Delta{pos_n}$ position difference. The spline weight $W_{spl}$ is finally obtained as

\begin{equation}
    W_{spl,i,j}(\Delta{pos_n}) = \sum_{d=1}^{(D+1)^2} Z_d(\Delta{pos_n}) \, W_{i,j} [idx_d(\Delta{pos_n})],
    \label{Eq_spl_conv_2}
\end{equation}

\noindent where $Z_d$ is a position-dependent bilinear-modulation coefficient associated with base $d$, and $idx_d$ is the position-dependent weight index attributed to base $d$. 

\begin{figure}[!t]
    \centering
    \includegraphics[width=\linewidth]{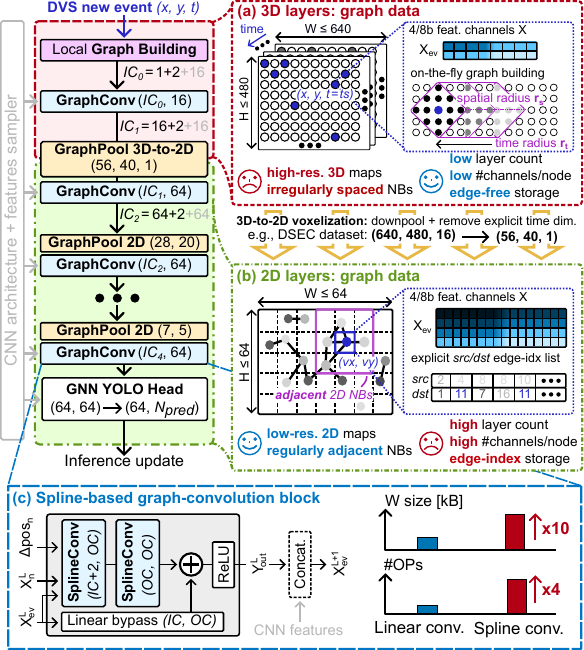}
    \caption{State-of-the-art DAGr-type workload (based on \cite{Gehrig2024}) and its key properties: (a) high-resolution 3D spatiotemporal layers, (b) low-resolution 2D spatial-only layers, and (c) graph-convolution blocks, with optional CNN-feature concatenation.}
    \label{Fig_algo}
    \vspace{-0.2cm}
\end{figure}




\subsection{From a Single Layer to a Scaled-up Network}
Modern EV-GNN workloads combine several graph-convolution layers into deep architectures tailored for high-resolution object-detection tasks, such as DSEC \cite{gehrig2021dsec}. The state-of-the-art DAGr-GNN network architecture (Fig.~\ref{Fig_algo}) has recently shown best-in-class accuracy on this problem, by alternating graph-convolution and graph-pooling layers that transform the graph structure, regularizing it and extending the spatial receptive field throughout the EV-GNN layers.

To that end, DAGr-GNN combines 3D and 2D layers with different core properties, which share a common graph-convolution kernel. The event-driven processing of a new event $(x, y, t)_{ev}$ goes as follows: first, the new event is fed to a graph-building layer, creating a local subgraph based on predefined spatial and temporal radii $r_s$ and $r_t$, respectively (Fig.~\ref{Fig_algo}(a)). This subgraph is then processed by a single \textit{3D spatiotemporal} graph convolution with only a few feature channels $X$ per node. Because 3D subgraphs are built on-the-fly, 3D edges do not need being stored. However, 3D layers still have to keep track over time of node features across the entire high-resolution map ($\leq 640\times480$ pixels), which are sparsely and irregularly distributed across space and time (Fig.~\ref{Fig_algo}(a)).
This 3D layer is directly followed by 3D-to-2D graph pooling, which removes the explicit time dimension and merges nodes spatially into different \textit{voxels}. This results in a lower-resolution ($\leq 64\times64$), spatial-only 2D voxel map (Fig.~\ref{Fig_algo}(b)) that encodes the explicit connectivity of 2D edges between voxels as a source-destination node-index list. Time information remains implicitly present through the evolution of this list over time. Moreover, potential neighbors in 2D maps are restricted to adjacent voxels, which regularizes the subgraph shape to a sparse 3$\times$3 square.
The 2D subgraph of the new-event's voxel $(vx, vy)_{ev}$ is then further processed and transformed by a series of 2D layers with a high number of feature channels $X$ per node (Fig.~\ref{Fig_algo}(b)). Eventually, it reaches a YOLO-like GNN head, which consists of six standalone 2D graph-convolution layers and outputs a prediction update.

At the core of DAGr-GNN, graph-convolution blocks are used to process feature data of both the 3D and 2D subgraphs. Shown in Fig.~\ref{Fig_algo}(c), these blocks combine two serial spline-convolution layers (SplineConv) with a linear-bypass layer. These blocks improve the overall task accuracy, at the cost of additional operations and storage requirements, notably due to the splines. Moreover, the layer's output can be optionally concatenated with features sampled from a CNN (e.g., ResNet-18), thereby combining the long-term structural information contained in processed RGB images with the fast, per-event update of the EV-GNN.

\begin{figure}[!t]
    \centering
    \includegraphics[width=\linewidth]{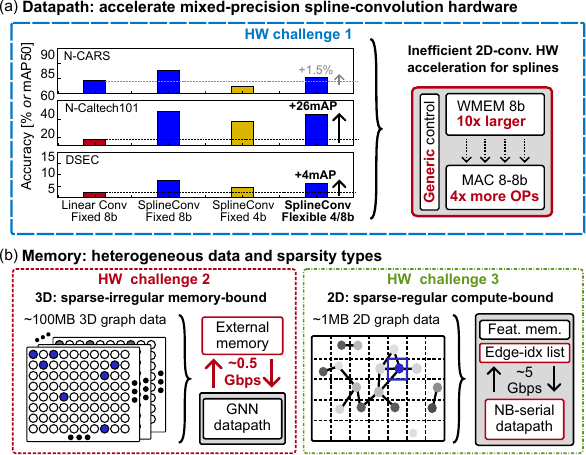}
    \caption{HW challenges stemming from the properties of state-of-the-art EV-GNN workloads on (a) the datapath and (b) the memory side.}
    \label{Fig_challenges}
\end{figure}

\subsection{From Algorithmic to Hardware Challenges}

The aforementioned properties of the DAGr-GNN network architecture induce a series of open HW challenges (Fig.~\ref{Fig_challenges}) for edge-vision processor chips.

On the \textbf{datapath side}, Fig.~\ref{Fig_challenges}(a) highlights the significant improvement in detection accuracy when using spline-based convolution over baseline, linear convolution in EV-GNNs, especially at iso 8b precision. Moreover, we observe that a mixed 4/8b precision per operand and per layer in DAGr-GNN enables significant accuracy benefits while limiting the storage and computational overhead of the splines, which entails a $10\times$ weight-memory and a $4\times$ operations/event increase at fixed 8b precision.
Yet, efficient HW support for spline convolution is missing in regular edge-vision accelerators, as regular MAC arrays and generic memory controllers (e.g., \cite{antonio2025open}) are ill-suited to the fast-yet-flexible processing of the dual MAC operation, the on-the-fly spline-index lookup and generation of the position-dependent $Z$ coefficients. These concerns yield a first HW challenge for EV-GNN processor chips to solve.

On the \textbf{memory side}, two additional HW challenges arise from the different properties of 3D and 2D layers, related to the sparsity and size of their graph data.
On the one hand, high-resolution 3D spatiotemporal maps require more than 100MB to store all possible graph-node features, which necessitates external-memory storage, e.g., in DRAM. Yet, the sparse and irregular nature of both the stream of new events and of the 3D spatiotemporal subgraphs leads to significant external-memory accesses (EMAs). This results in a memory-bound 3D regime and a second HW challenge. 
On the other hand, low-resolution 2D spatial-only maps can be fully stored on-chip within 1MB, with a sparse but regular graph structure. However, the source-destination list of 2D edges also has to be stored for these 2D layers, in order to keep track of the graph structure. Accesses to indices stored in this list correspond to irregular memory fetches, which cannot be easily parallelized and favor a neighbor-serial dataflow, similar to \cite{Yang2025}. The low parallelism of this dataflow limits the peak throughput of the system, resulting in a compute-bound 2D regime and a third HW challenge. 

\section{Chip Architecture}
\label{sect:architecture}

\begin{figure}[t!]
    \centering
    \includegraphics[width=\linewidth]{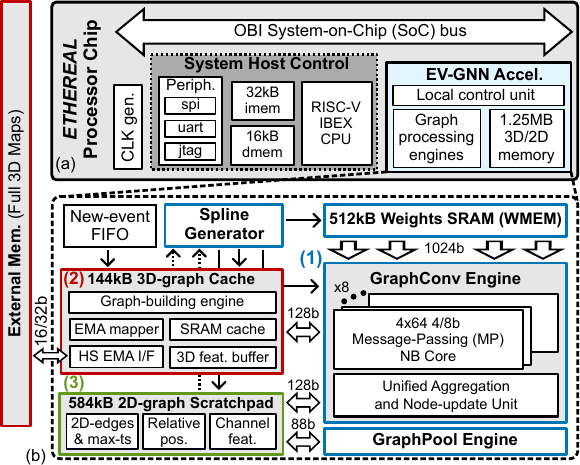}
    \caption{Overview of (a) the ETHEREAL chip and (b) the EV-GNN accelerator.}
    \label{Fig_arch_overview}
\end{figure}

To solve these challenges, we propose the ETHEREAL processor chip shown in Fig.~\ref{Fig_arch_overview}(a). Built atop an MCU baseline \cite{sauter2025croc} with a dedicated RISC-V-based CPU host for control and an open-bus interface (OBI) system bus, ETHEREAL embeds a loosely coupled EV-GNN accelerator with local configuration, graph-processing engines and a 1.25MB 3D/2D-split memory that can either be directly accessed by the accelerator, or by the CPU.
The accelerator's top-level block diagram is shown in Fig.~\ref{Fig_arch_overview}(b), excluding control units. We propose three key features to address our three challenges.

First, we introduce (1) a dedicated spline-generator unit and a graph-convolution engine tightly coupled to a 512kB weight memory (WMEM) to solve our first HW challenge. These blocks provide support for both linear and spline convolutions, and offer a flexible 4b or 8b resolution per MAC operand, which can be adapted per layer at runtime. The graph-convolution engine consists of eight neighbor-parallel message-passing (MP) cores and a unified aggregation and node-update unit. The spline generator handles the generation of spline indices and bilinear coefficients, while indexed weights $W$ are directly provided to the datapath through a 1024b direct-streaming bus. In addition, a graph-pooling engine accelerates the max-feature, mean-position and directed-edge pooling operations described in \cite{Gehrig2024}.

Second, we introduce (2) a 144kB 3D-graph cache memory combined with the graph-building engine to address our second HW challenge. The 3D cache exploits the locality of events within spatiotemporal regions of interest, where nearby pixels are more likely to spike together and to serve as subgraph neighbors to new events. We therefore keep the features of these more regularly accessed nodes locally on chip, thereby significantly reducing the amount of slow and energy-hungry EMAs. This 3D memory contains the SRAM cache itself, a 3D-reshaping output buffer that matches the datapath bandwidth, an EMA mapper that handles off-chip address generation and a high-speed (HS) interface for external transfers at 0.25-0.5Gbps, which uses a custom bit-parallel SPI-like protocol to emulate EMAs to the full 3D off-chip maps.

Third, we propose (3) a 584kB 2D-graph scratchpad memory to exploit the sparse yet regular $K\times K$ structure of 2D neighborhoods to provide graph data to the engines with a high bandwidth, enabling in turn a high utilization of the neighbor-parallel cores without any lookup latency penalty to solve our third HW challenge. The 2D memory consists of three blocks: a first to efficiently store the 2D edges and per-voxel timestamp of the 2D maps, which avoid the source-destination look-up cost per neighbor; a second for the relative position $(rx, ry)$ of nodes within each voxel and the total event count associated to that voxel; and a third for the voxels' features. The 2D scratchpad provides a direct, high-bandwidth read/write interface to both the graph-convolution and graph-pooling engines for a full on-chip streaming of 2D graph data, thereby maximizing the peak utilization of parallel processing elements in the graph-processing engines.

The host-accelerator communication goes as follows: after configuring the accelerator, the host CPU triggers the execution of one layer and waits for an IRQ to indicate the end of the accelerator's operations. In that context, a layer can either be a graph-convolution block, a standalone graph-convolution, graph-building or graph-pooling operation, or a mixed sequence of these three operations. Scheduling optimizations are later discussed in Section V.C.

\section{Graph-convolution Engine and Dataflow}
\label{sect:datapath}


Our first HW challenge relates to the efficient execution of splines, which we enable by revisiting the spline-convolution dataflow to enable a higher level of parallelism. In what follows, we start by discussing this dataflow and its execution in hardware using a dedicated spline generator (Section~\ref{sect:datapath}.A). We then present the graph-convolution engine that exploits the dataflow properties to minimize the per-layer execution time and energy (Section~\ref{sect:datapath}.B-C).

\subsection{Spline-convolution Dataflow}
To unlock \textit{neighbor-level parallelism} and thereby increase our system's throughput, we propose to leverage a \textit{spline-iterative} formulation of the message-passing algorithm described in Section II.A. To that end, we swap the order of the two MAC operations in Eqs. (\ref{Eq_spl_conv_1}) and (\ref{Eq_spl_conv_2}), yielding

\begin{equation}
    Y_{msg, n, j} = \sum_{d = 1}^{(D+1)^2} Z_d(\Delta{pos_n}) \; Y_{MAC, n, j}(\Delta{pos_n}),
    \label{Eq_msg_1}
\end{equation}

where

\begin{equation}
    Y_{MAC, n, j}(\Delta{pos_n}) = \sum_{i = 0}^{IC-1} X_{n, i} \; W_{i,j}[idx\textcolor{gray}{(\Delta{pos_n})}].
    \label{Eq_msg_2}
\end{equation}

\begin{figure}[!t]
    \centering
    \includegraphics[width=\linewidth]{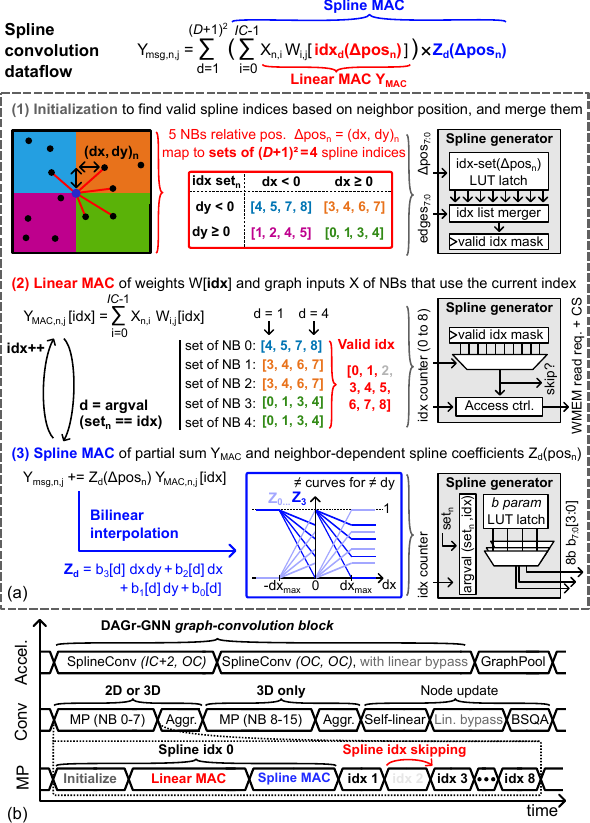}
    \caption{(a) Three-phase decomposition of the spline-based MP operation, controlled by the accelerator's spline generator. (b) Illustration of the accelerator, graph-convolution and message-passing dataflows.}
    \label{Fig_dataflow}
\end{figure}

\noindent By applying $Z$ coefficients once per index to the 20b partial sum $Y_{MAC}$ (Eq. (\ref{Eq_msg_1})) instead of applying $X$ per channel to an 18b $W_{spl}$ (Eq. (\ref{Eq_spl_conv_1})), this formulation reduces the number of high-precision MACs by a factor $IC/(D+1)^2 \gg 1$. 
We further decompose these expressions into a three-phase dataflow, in which we parallelize computations over both neighbors and output channels $OC$. The generation of the required spline variables is handled by the \textit{spline generator} unit, as illustrated in Fig.~\ref{Fig_dataflow}(a) for a five-neighbor example.

First, (1) an \textbf{initialization phase} determines the unique set of valid spline indices shared by the neighbors of the new event node. Interestingly, we observe that \textit{sets} of $(D+1)^2 = 4$ indices are fixed per spatial quadrant around the event's node. This property stems from the finite nature of the spatial radius between an event and its potential neighbors, both in 2D and 3D layers. We exploit this observation by storing the index sets in configuration lookup tables (LUTs), and by looking up the set associated with each neighbor during initialization, based on $\Delta{pos_n}$. A merger can then determine the $K \times K = 9$b combined one-hot mask of \textit{valid} spline indices, based on the edge connectivity. In the example case, indices from the blue, orange and green regions are merged together, while index 2 is left out, as it is only present in the set of the neighbor-less purple region.

Then, (2) a \textbf{linear-MAC phase} computes the partial sum $Y_{MAC}$ between the layer's neighbor features $X_n$ and weights $W[idx]$ at the current index counter (reset to 0). By comparing this counter to the valid index mask, the spline generator determines whether the index should be processed or can be skipped (e.g., $idx = 2$), a mechanism we call \textit{spline skipping}. If the index is deemed valid, the generator outputs a read request to the weights memory at the target index, and generates control signals (CS) to enable the involved MP cores and (3D or 2D) graph-data memories, as depicted in Fig.~\ref{Fig_arch_overview}(b). The fine-grained enabling of the different MP cores as a function of the subgraph's neighborhood is discussed below.

Finally, (3) a \textbf{spline-MAC phase} between the partial sum $Y_{MAC}$ and the bilinear spline coefficient $Z_d$ contributes to the accumulation of messages $Y_{msg}$, for all neighbors $n$ relevant to the current index. The spline coefficient is obtained from a clipped bilinear interpolation, whose four parameters $b_d$ are precompiled and stored into configuration LUTs. The set of $b$ parameters to apply to each neighbor changes with the position-dependent set, but also with the location within the set, such that $d = \mathrm{argval}(set_n == idx)$. For instance, when $idx = 7$, $d = 3$ for neighbor 0 (blue), $d = 4$ for neighbors 1-2 (orange), and no accumulation occurs for neighbors 3-4 (green), which do not involve this index value.

Phases (2) and (3) are repeated until all valid indices have been iteratively covered, after which the aggregation phase takes place (Fig.~\ref{Fig_dataflow}(b)). 2D layers then directly jump to the node-update phase, whereas 3D layers could be repeated a second time in cases where the 3D neighborhood exceeds 8 neighbors. Graph pooling directly follows the node update when required, preventing the storage of pre-pooled graph-feature maps and finishing the processing of a standalone graph-convolution layer. For a graph-convolution block, the same set of operations apply directly twice in a row without CPU intervention. In that case, an additional linear transformation occurs during the node-update phase to add the contribution of the linear-bypass layer of the block (Fig.~\ref{Fig_algo}(c)).

\begin{figure}[!t]
    \centering
    \includegraphics[width=\linewidth]{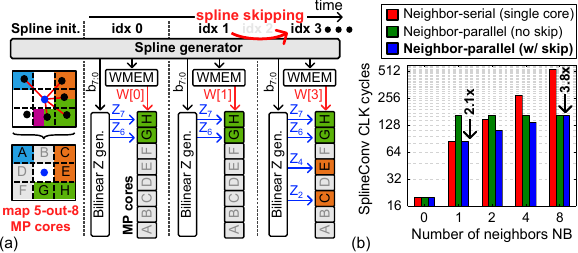}
    \caption{(a) High-level example of the spline-based message-passing operation in the EV-GNN accelerator. (b) Comparison of the total graph-convolution duration for different dataflows (at fixed memory bandwidth).}
    \label{Fig_perf_spl_conv}
\end{figure}

With this spline-iterative dataflow, each large 2D weight tensor $W[idx]$ is fetched only once and consumed in parallel by all neighbors that use it. This \textit{neighbor-level parallelism} enhances the overall throughput and avoids repeated energy-hungry accesses to the 512kB WMEM. 
The high-level execution of the message-passing operation in the EV-GNN accelerator is illustrated in Fig.~\ref{Fig_perf_spl_conv}(a), for the example case in Fig.~\ref{Fig_dataflow}. Neighbor-level parallelism is achieved by mapping the processing of each of the five neighbors to a different MP core, labeled from A to H. After initialization, valid spline indices are sequentially processed, whereas unused ones are \textit{skipped}. At each index, 1024b channel data of the 2D weight tensor $W[idx]$ are fetched per clock cycle and broadcast to all active neighbor cores in parallel. To save energy, the spline generator clock- and data-gates unused cores per screened index.
Compared to a neighbor-serial approach with a single core at fixed WMEM bandwidth, this neighbor-parallel processing improves the end-to-end execution time of the entire graph-convolution process by up to 3.8$\times$ on average (Fig.~\ref{Fig_perf_spl_conv}(b)). Moreover, spline skipping proves to be key in preserving performance at low neighbor count under neighbor-parallel processing.

\begin{figure}[!t]
    \centering
    \includegraphics[width=\linewidth]{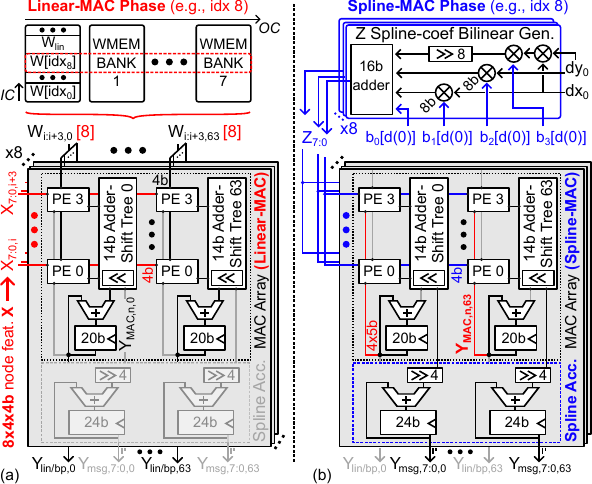}
    \caption{Message-passing core's operation during (a) linear-MAC and (b) spline-MAC phases of the spline-based graph-convolution.}
    \label{Fig_datapath}
\end{figure}

\subsection{Dual-mode Message-passing Core}
To support this dataflow, the graph-convolution engine embeds eight dual-mode MP cores specialized for spline-based operations. Each core consists of a reconfigurable MAC array with 4$\times$64 5/4b processing elements (PEs) followed by spline accumulators, as detailed in Figs. \ref{Fig_datapath} and \ref{Fig_PE}. The cores are output stationary and keep all partial sums locally in registers during the message-passing operation.
To provide a flexible trade-off between accuracy, storage and speed/energy, we support both 4b and 8b precision for the input and weight operands of the MAC array, which can be changed per layer at runtime. In this way, we turn the low sensitivity of most EV-GNN layers to 4b quantization on edge-vision datasets, as seen back in Fig.~\ref{Fig_challenges}(a), into actual performance benefits.

The MAC array operates in two modes, which correspond to phases (2) and (3) of the spline-iterative processing.

\vspace{0.2cm}

\subsubsection{Linear-MAC phase (Fig.~\ref{Fig_datapath}(a))} (un)signed $8 \times 4 \times 4$b input features $X$ are broadcast over the columns of the different MP cores, which compute the message of a different neighbor each. These inputs feed the first 4b entries of the PEs, whereas their second entry is fed by $4 \times 64 \times 4$b (un)signed weights $W[idx]$, shared across all cores. The MAC result is accumulated in a 20b output register, yielding partial sums $Y_{MAC}$ after $IC/4$ cycles. When required, two additional cycles serve to add the contribution of the 2$\times$8b absolute position difference $(|dx|, |dy|)$ to the message.

\begin{figure}[!t]
    \centering
    \includegraphics[width=\linewidth]{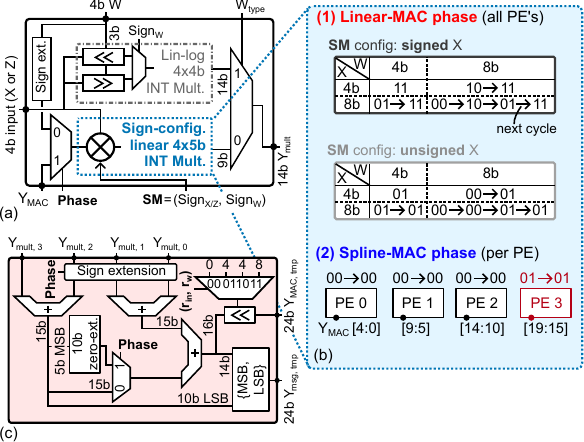}
    \caption{(a) Reconfigurable PE architecture, with (b) details of its configuration modes. (c) Mode-dependent adder tree, with bit alignment for multi-precision support.}
    \label{Fig_PE}
\end{figure}

To support operands with different bit precisions and/or distribution types on a per-layer basis, we propose configurable PEs that embed both a 4$\times$4b lin-log shift multiplier as well as a 4$\times$5b sign-configurable linear multiplier (Fig.~\ref{Fig_PE}(a)).
The former is tailored for layers with a wide distribution of weight values, and is restricted to be used with 4b/4b input/weight operands.
The latter accommodates for a flexible 4b-or-8b precision per operand, which is supported by means of a 1/2/4 multi-cycle execution with 8b data split into 4b LSB/MSB parts that are sequentially fed to the engine. 


\vspace{0.2cm}

\subsubsection{Spline-MAC phase (Fig.~\ref{Fig_datapath}(b))} unsigned $Z$ coefficients are fed as first PE operand instead of the input features. These coefficients are generated by a bilinear generator based on the selected, position-dependent $b$ parameters (see Section III.B). Simultaneously,  the second PE entry is fed with different 5b slices of the partial sum $Y_{MAC}$, with the 5b MSBs fed to PE 3 as a signed entry, and the 3$\times$5b LSBs fed as unsigned entries to PEs 2-0, exploiting the PE's per-row sign-configurability (Fig.~\ref{Fig_PE}(b)). 
Moreover, the inner sum of the adder tree (Fig.~\ref{Fig_PE}(c)) is modified by additional muxing, zero-extension and final concatenation to properly align the contributions of the different PEs. This adder-tree output is finally added to the core's message $Y_{msg}$ for every output channel in parallel. 
Interestingly, the area overhead of the MAC-array part remains below 30\% compared to a baseline MAC array with similar input-operand bit precisions, thereby enabling an efficient acceleration of spline convolution with minimum overheads.

\subsection{Unified Aggregation-and-node-update Unit}
\begin{figure}[!t]
    \centering
    \includegraphics[width=\linewidth]{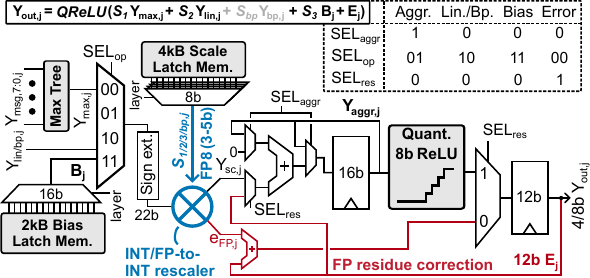}
    \caption{Architecture of a unified aggregation and node-update unit, highlighting its mixed INT/FP-to-INT rescaler with an error-correction logic for the accumulated FP-to-INT conversion residue.}
    \label{Fig_BSQA}
\end{figure}

The output message $Y_{msg}$ of all eight MP cores as well as the linear MAC result $Y_{MAC}$ of MP core 0 are sent to a unified aggregation-and-node-update unit (Fig.~\ref{Fig_BSQA}), for all 64 columns in parallel. It consists of a configurable datapath with a max-tree logic, two phase-dependent 16b and 12b accumulators, respectively, and a mixed INT/FP-to-INT rescaler. The latter takes a 22b integer input and an FP8 scale factor $S$ with a 3b mantissa and a 5b exponent as inputs, and produces a 16b rescaled output $Y_{sc}$ and a 10b rounding-error residue term $e_{FP}$. The unit yields the convolution output $Y_{out}$ as a sum of rescaled contributions:

\begin{equation}
    Y_{out} = QReLU(S_1 \, Y_{max} + S_2 \, Y_{lin} \textcolor{gray}{+ S_{bp} \, Y_{bp}} +  S_3 \, B + E ),
\end{equation}

\noindent where $Y_{max}$ is the per-channel max-feature aggregation of the eight MP-core messages $Y_{msg, 7:0}$, $Y_{lin}$ (resp. $Y_{bp}$) is the self-linear (resp. optional linear-bypass) term described in Fig.~\ref{Fig_algo}, $B$ is the column-wise bias term and $E$ is the accumulated 6b rounding error from the FP-to-INT conversion over prior cycles. 
During the self-linear phase (resp. linear-bypass phase), MP core 0 is used to yield a linear MAC result $Y_{lin}$ (resp. $Y_{bp}$) from the graph-convolution layer's (resp. block's) input event features $X_{ev}$, whereas other cores remain clock- and data-gated.
Moreover, the unit keeps track of the accumulated FP-to-INT conversion-error residue $E$ over the full convolution, and compensates for it in a last output-accumulation phase. By representing $E$ on 12b, the 3-$\sigma$ rounding error can be contained within $\pm$1 LSB, thereby avoiding the accumulation and eventual divergence of the error across consecutive events.
Finally, the corrected 16b-aggregated output undergoes a 4b or 8b quantized-ReLU ($QReLU$) activation, whose result is stored into the 8 LSBs of the 12b output register. This output is then either written to the 3D/2D memory as a new graph-node state $X$, or directly forwarded to the pooling engine.

\section{Memory Hierarchy}
\label{sect:arch}

To address the remaining memory-bound and compute-bound regimes of state-of-the-art EV-GNN workloads, respectively tied to the irregular and regular sparsity of 3D and 2D graph-data maps, we propose to split the accelerator memory in two specialized parts. After addressing each part individually, we also showcase how to optimize the inter-layer execution scheduling to minimize the end-to-end latency per event inference. 

\subsection{3D Spatiotemporal Cache Memory}
\begin{figure}[!t]
    \centering
    \includegraphics[width=\linewidth]{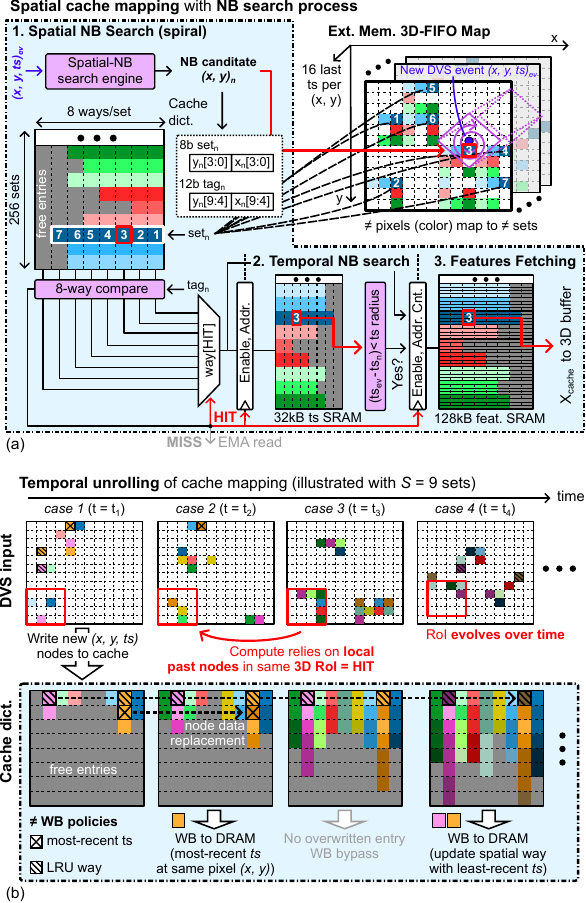}
    \caption{Principle and architecture of the 3D spatiotemporal cache with intertwined graph building, with illustration of (a) spatial mapping and impact on the neighbor-search process, (b) temporal evolution of the cache content and write-back conditions.}
    \label{Fig_cache_1}
\end{figure}

\begin{figure}[!t]
    \centering
    \includegraphics[width=\linewidth]{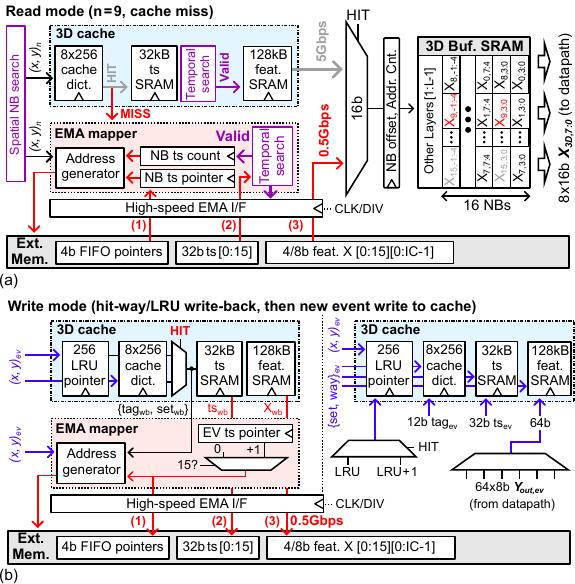}
    \caption{Overall architecture of the 3D data memory subsystem and its direct-accessed external memory, under (a) neighbor-search read and (b) event write after write-back operation. The sequential order of accesses to external memory sections, when required, is pointed out.}
    \label{Fig_cache_2}
\end{figure}

We first introduce a 3D cache architecture (Figs. \ref{Fig_cache_1} and \ref{Fig_cache_2}) to leverage the spatiotemporal locality in the EV-GNN 3D maps. This locality stems from spatiotemporal regions of interest (RoIs) in the DVS 3D input, which are more likely to generate events than others during a lapse of time. Typically, these RoIs correspond to moving objects and only occupy a small fraction of the entire DVS receptive field, in particular in setups with a static DVS camera \cite{Verma2024}. 
Our 3D cache exploits the inherent existence of RoIs by storing the position and 3D features of new-event nodes as soon as they are processed, thereby keeping track of the evolving RoIs on the chip.

To balance hardware overhead and cache hit rate, we propose a set-associative cache with 256 sets, 8 spatial and 1 temporal ways, intertwined with the graph building process. The \textit{spatial} mapping and neighbor-search process is illustrated in Fig.~\ref{Fig_cache_1}(a): upon a new DVS input event, a spatial-then-temporal search for neighbor candidates is initiated. First, the spatial position $(x, y)_n$ of the next neighbor candidate is derived from a spiral-shaped L1 search \cite{Yang2025}. After extracting tag and set information from this position, a cache dictionary lookup takes place: upon a hit, a temporal-search process compares the difference $dt = ts_{ev} - ts_n$ to a target temporal radius $r_t$, with $ts_{ev}$ and $ts_{n}$ respectively the event's and $n$-th neighbor's timestamp. 
If $dt$ lies within radius, the neighbor candidate is accepted, and its cached 3D features are transferred from the 128kB features SRAM to a 3D reshaping buffer. Upon a miss, the search process looks for valid neighbor candidates in the external memory at the target $(x, y)_n$ location following a three-step process, described in Fig.~\ref{Fig_cache_2}(a): first (1), the offset address of the most-recent event for the $(x, y)_n$ is read from the external memory; then (2), the corresponding 32b timestamp of the node is retrieved for a temporal-search validity check; finally (3), if the timestamp is valid, the node features of the neighbor candidate are transferred to the 3D reshaping buffer.
This process repeats until 16 neighbors have been found, or until the spiral search reaches a maximum configurable radius $r_s$.
Then, 3D graph-convolution operations start as described in Section IV, fetching 128b data from the 3D reshaping buffer per cycle and distributing them across the eight MP cores. To match this bandwidth, write transfers to the buffer from either the cache or the external memory reshape the channel-first encoding of cached features in a neighbor-first format, with $8\times4\times4$b features per buffer row (Fig.~\ref{Fig_cache_2}(a)). This reshaping allows fully parallelizing data fetches during graph convolution, while keeping the structure of the cache dictionary simple. 

\begin{figure}[!t]
    \centering
    \includegraphics[width=\linewidth]{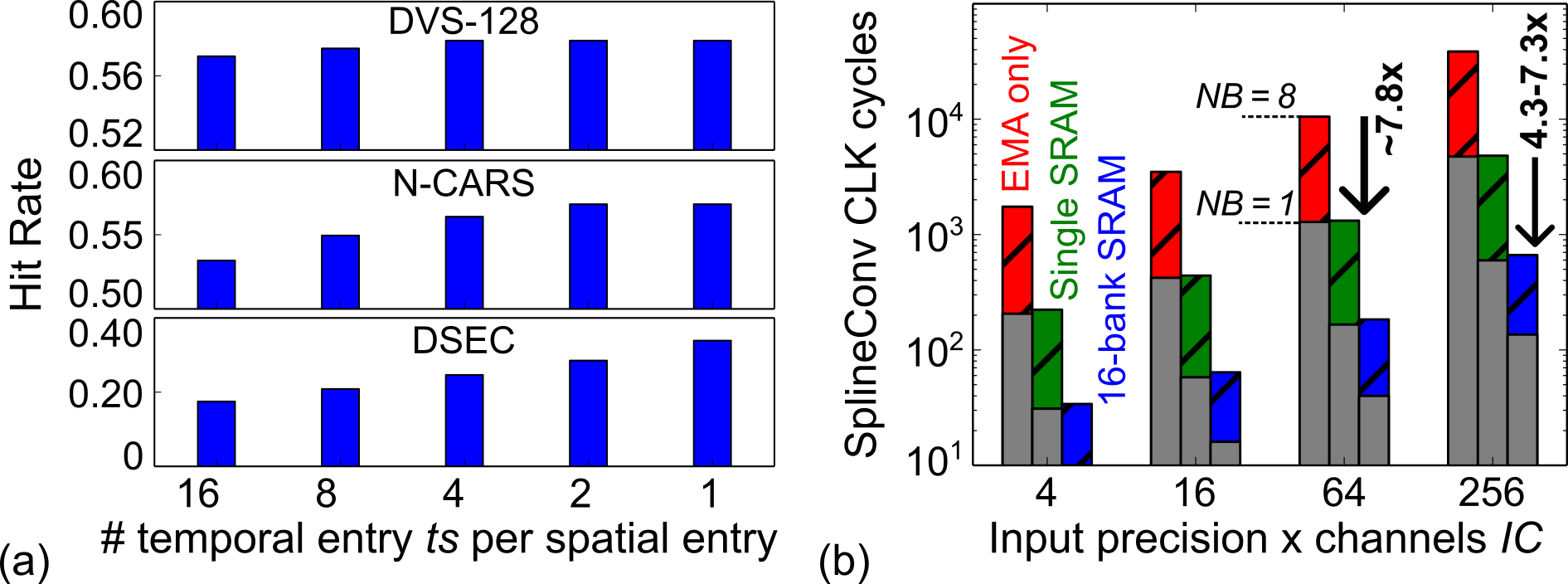}
    \caption{Performance benefits of the split 3D/2D memory hierarchy. (a) Hit rate for different 3D spatiotemporal cache lookup strategies, at fixed number of sets. (b) Comparison of spline-convolution latency with external-memory, single-SRAM and interleaved banked-SRAM 2D graph-data storage.}
    \label{Fig_mem_perf}
\end{figure}

To maximize cache hits over time, the 3D cache follows a most-recent-timestamp policy where only the most-recent event per $(x, y)$ entry can be stored in the cache at any time. This policy choice is based on experiments carried out on different edge-vision datasets at a fixed number of cache entries and events, as evaluated in Fig.~\ref{Fig_mem_perf}(a). Consequently, EMAs can either take place upon a miss, as described above, or after a hit with a valid timestamp found in the cache: in this case, other neighbors at the same $(x, y)$ location might be present in the external memory, and are sequentially looked for using a same EMA pattern from the second-most-recent one.

The resulting evolution of the cache content over time is illustrated in Fig.~\ref{Fig_cache_1}(b), across four consecutive timestamps (case 1-4). In ETHEREAL, the features of all 3D layers associated to a new event $(x, y,t)_{ev}$ are automatically written to the cache as soon as they are computed. For the DAGr-GNN workload, this amounts to 32 bytes per cache entry. This mechanism ensures the on-chip availability of the new event's features during 3D self-linear and linear-bypass operations, which boosts the hit rate by 10 to 15\%. 
Consequently, node-data replacement in the cache can happen in two scenarios. On the one hand (case 2), when a given $(x, y)_{ev}$ entry is already present in the dictionary, the cached node's timestamp $ts$ and features $X$ are replaced by the data of the new event, thereby following the most-recent-timestamp mapping policy. On the other hand (case 4), when the target cache set is full, the least-recent-updated (LRU) way per set is replaced. In both cases, the replaced cache data undergo a write-back (WB) procedure to the external memory, unless the past node is outdated. The WB procedure takes place along three phases that mirror the EMA neighbor-search process, as shown in Fig.~\ref{Fig_cache_2}(b).

Altogether, the 3D cache's 30-to-60\% hit rate turns into a 1.3-to-2.3$\times$ reduction in total EMAs, thereby alleviating the second HW challenge of EV-GNNs. To further mask the latency of 3D layers, two options could be considered. A first option would be to increase the hit rate by means of a higher degree of associativity, a larger cache size and/or a dynamic cache-allocation policy. These changes, however, would bring significant HW overheads and an increased design complexity. Instead, we propose to hide the remaining EMA latency by parallelizing 3D and 2D operations in DAGr-type workloads through overlapping scheduling, a solution discussed in Section V.C.


\subsection{2D Spatial Scratchpad Memory}
We also introduce a 2D graph-data scratchpad for the full on-chip storage of 2D voxel maps.
To enable a latency-free, high-bandwidth streaming of graph data to the different computing engines, the 2D memory exploits the adjacency property of 2D neighborhoods in two ways.

\begin{figure}[!t]
    \centering
    \includegraphics[width=\linewidth]{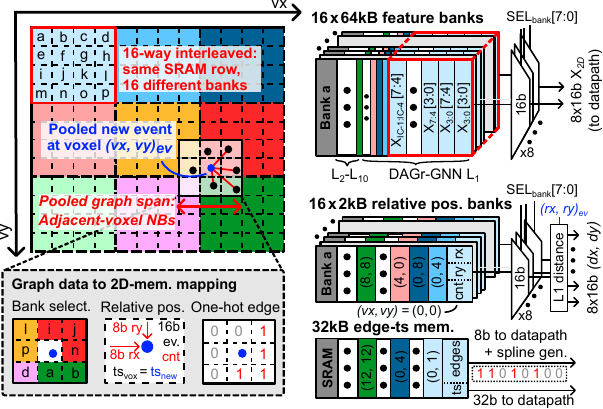}
    \caption{2D graph-data encoding and mapping to different memory banks.}
    \label{Fig_spad}
\end{figure}

On the one hand, to provide a continuous stream of input data to the eight MP cores during graph convolution, we map the feature and position data of the different 2D voxels to 16 banks each. This mapping follows the 2D-interleaved pattern depicted in Fig.~\ref{Fig_spad}: in this way, up to 8$\times$4$\times$4b features $X_{2D}$ or position differences $(dx, dy)_{7:0} = (rx, ry)_{7:0} - (rx, ry)_{ev}$ can be fetched from eight different memory banks without any stall. The addresses of these banks, and that of the neighboring voxel rows to access, is entirely derived from the current event's voxel $(vx, vy)_{ev}$ by the 2D memory controller.
Feature banks store 4$\times$4b channels per row, with 4b MSB/LSB slices split in two consecutive rows when utilizing 8b features. This split matches with the datapath's serial multi-bit compute described in Section IV.B. The number of rows allocated to a voxel thus depends on the channel depth and precision of a layer's features, and can be changed at runtime.
In contrast, relative-position banks always store the voxel's node 2$\times$8b relative position $(rx, ry)$ in a single row, together with the 16b event counter of the voxel. This number is used during graph-pooling for computing the updated voxel's relative position through an accumulated mean averaging.

On the other hand, to avoid slow inter-layer graph-building processes, the 2D memory explicitly keeps track of the graph structure of the lower-resolution 2D maps by storing their edge connectivity. Furthermore, it avoids the lookup latency of source-destination list by encoding the 8b 2D-edge and 32b timestamp data of each voxel as one row of a single SRAM bank. Notably, we leverage the directionality of 2D edges to one-hot encode them on 8b, so that each bit indicates the existence (bit-1) of a valid inbound edge from one of the eight adjacent voxels. During graph convolution, this 8b vector is sent together with the relative position $(dx, dy)$ of the eight node's neighbors to the spline generator, to determine the MP core mask in 2D mode. During graph pooling, the edge and timestamp information of both the current event's voxel and its adjacent voxels is sent to the graph-pooling engine to undergo an update, including new-edge generation and edge inversion \cite{Gehrig2024}.

Altogether, our 2D graph-data memory enables neighbor-parallel computing without any preprocessing or graph-building latency penalty, thereby being the key ingredient to overcome the compute-bound regime of our third HW challenge. As shown in Fig.~\ref{Fig_mem_perf}(b), this mapping achieves a 4.3-to-7.3$\times$ latency reduction (in clock cycles) compared to a single bank with a neighbor-serial datapath, with higher gains obtained in the presence of denser neighborhoods. Compared to an EMA-only storage with a 0.5Gbps bandwidth, similar to our high-speed interface (Section~\ref{sect:architecture}), the per-layer execution time is reduced by up to 57$\times$. Further gains could be leveraged by transferring more channels at once, but requires more PEs to consume these data, as well as a larger weight-memory bandwidth.

\subsection{System-level Optimization}
\begin{figure}[t!]
    \centering
    \includegraphics[width=\linewidth]{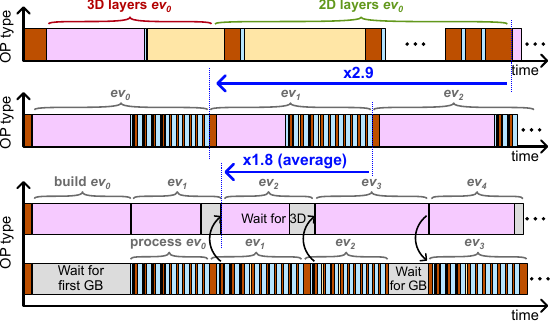}
    \caption{Qualitative illustration of system-level operation scheduling. (a) Baseline serial execution with CPU pooling and per-layer full reconfiguration, (b) serial execution with accelerated pooling and multi-layer precompiled configuration, (c) parallel graph-building and graph-processing execution, enabled by the decoupling between 2D and 3D memories.}
    \label{Fig_scheduling}
\end{figure}

With all three HW challenges addressed on the graph-convolution side, we can take a step back and look at ETHEREAL's system performance on the DAGr-GNN workload. Fig.~\ref{Fig_scheduling}(a) depicts a baseline breakdown of the different tasks and layers performed serially, with only the graph-convolution part being accelerated. However, the end-to-end latency per event then becomes fully dominated by the graph-building process, the graph-pooling operations, and the full inter-layer reconfiguration of the accelerator by the CPU, thereby hiding the benefits of the graph-convolution engine. RTL-level simulations showcase that a single event is processed end-to-end in 100-to-150$\mu$s under such conditions, thereby failing to achieve the low-latency target.

To overcome these limitations, we first propose to accelerate graph pooling with a dedicated engine (Section III) and to embed an array of configuration registers inside the accelerator's controller. The latter allows storing the configuration of eight layers at a time, such that the CPU only has to handle context-switching and configuration-selection between layers. The impact of these combined techniques is shown in Fig.~\ref{Fig_scheduling}(b), where the end-to-end latency per event is reduced by 2.9$\times$ compared to Fig.~\ref{Fig_scheduling}(a) on average.

Finally, we leverage the split between 3D and 2D memories to deal with the remaining graph-building bottleneck, by pipelining the graph-building operation of the \textit{next event} $ev_{k+1}$ (3D memory) with the processing of the 2D layers of the \textit{current event} $ev_k$ (2D memory + processing engines). As these phases roughly take the same total time, the graph-building latency can be mostly hidden at the system level, resulting in an additional improvement of the latency per end-to-end event-wise inference by 1.8$\times$. 
Nonetheless, the respective durations of the graph-building and graph-processing operations still vary on an event-per-event basis, thereby requiring cross-engine synchronization. To that end, each engine sends an IRQ to the CPU upon task completion. When both IRQs have been received, the CPU triggers the processing of the 3D layers of the next event, before launching 2D graph-convolution operations together with the graph building of the next event.

\section{Measurement Results}
\label{sect:arch}

The 3.7mm$^2$ ETHEREAL processor chip was fabricated in TSMC 28nm bulk CMOS. The chip's microphotograph in Fig.~\ref{Fig_meas_setup}(a) highlights the placement of SRAM macro banks on each side of the graph-convolution engine's datapath: WMEM banks outputs can be directly broadcast to the different MP cores, whereas a distributed mux-tree logic selects the feature and position banks, as well as the 3D or 2D memory source. 
To pass timing closure at 250MHz during physical implementation, a mix of LVT and SVT devices were used for the logic, whereas HVT SRAM periphery was adopted to limit the total leakage below 10mW. In order to pass hold fixing in the FF 125$^\circ$C corner during routing, HVT devices were also used for clock-tree buffers after the clock-tree synthesis, resulting in a final chip density of 71.5\%. Experiments at design time showed that other combinations of $V_t$ types would not lead to a successful design, either because of a too high routing congestion, or due to a leakage-power increase beyond 10\% of the chip's active power, which was not deemed acceptable for an edge-device target.

\subsection{Experimental Setup}
\begin{figure}[!t]
    \centering
    \includegraphics[width=\linewidth]{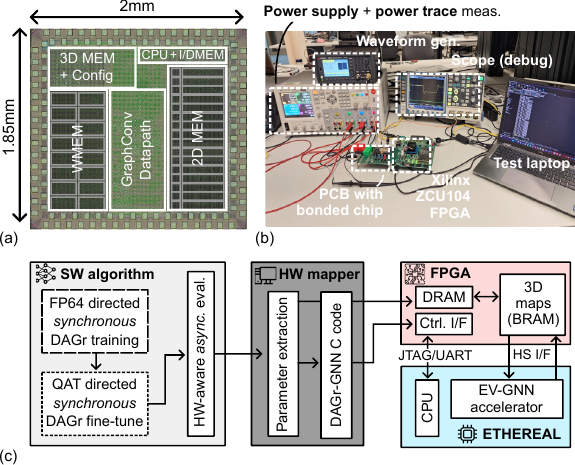}
    \caption{(a) Chip microphotograph, (b) measurement setup, and (c) algorithm-to-hardware deployment flow.}
    \label{Fig_meas_setup}
\end{figure}

The lab measurement setup is shown in Fig.~\ref{Fig_meas_setup}(b). The bonded chip is plugged into a PGA socket on a custom PCB, which connects both its low-speed control (JTAG/UART/SPI) and high-speed data interface (HS I/F) through an FMC connector to a Xilinx ZCU104 FPGA board. The 1.8V I/O and nominal 0.9V logic and memory power supplies are all provided externally. Lastly, a test laptop serves as an interface to the FPGA's running host, which handles the programming, data transfers, and debugging of the chip.

The full SW-to-HW deployment pipeline for the DAGr-GNN workload is shown in Fig.~\ref{Fig_meas_setup}(c). After full-precision and quantization-aware training (QAT) of the network, a modeled asynchronous evaluation with HW-aware parameters takes place to evaluate the expected chip output. A mapper then extracts and formats the necessary parameters, which can then be used by the C-code implementation of the DAGr-GNN algorithm for ETHEREAL. At programming time, the compiled C-code is transferred to the chip through the FPGA's control interface. The FPGA also stores the full 3D maps, serving as an external-memory emulator. To simplify transfers, a local block RAM (BRAM) stores the full 3D map of the event-stream sample being processed, to be accessed by the EV-GNN accelerator's HS I/F interface, as well as the expected output of every layer from the HW-aware EV-GNN algorithmic model.
For each sample, a stream of 10$^3$ new events is stored in the chip's DMEM. ETHEREAL processes one event at a time, and sends the final results back through SPI. We evaluate in this way the entire DAGr-GNN architecture on-chip, except for its three last layers (within the YOLO head), which need 32b representation for correct mAP computation. The results of these layers are sent back to the FPGA after the processing of the full $10^3$ event stream, where they are compared with the expected results for validation. Execution time and power measurement is performed by analyzing current traces with a power analyzer, identifying the rise and fall of the traces between the chip's $\sim$10mW standby regime and its $\sim$100mW processing regime.

\subsection{Electrical Characterization}
\begin{figure}[!t]
    \centering
    \includegraphics[width=\linewidth]{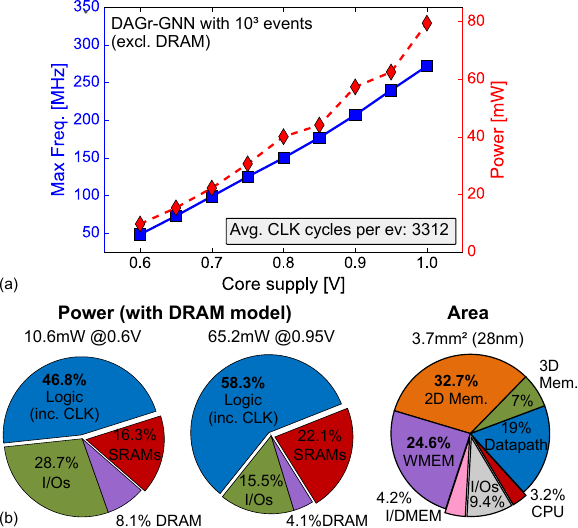}
    \caption{(a) Voltage-frequency-power curve of the chip, measured on DAGr-GNN for 10$^3$ random events. (b) Power and area breakdowns.}
    \label{Fig_elec_charac}
\end{figure}

We first report the electrical characterization of ETHEREAL in Fig.~\ref{Fig_elec_charac}, averaged on a stream of 10$^3$ events on the DAGr-GNN workload, using synthetic data that follow the DSEC graph-data distribution. ETHEREAL's maximum system-level frequency ranges from 50 to 280MHz when sweeping the logic and memory supply voltages from 0.6V to 1V (Fig.~\ref{Fig_elec_charac}(a)). 
During DAGr-GNN end-to-end evaluation, the total chip power (including a DRAM model based on CACTI \cite{cacti_2008} for the EMA energy) increases from 10.6mW (0.6V) to 65.2mW (0.95V, nominal) in nonlinear steps, related to stepwise changes of clock-division factor of the high-speed interface's clock to preserve functionality, at the cost of additional end-to-end inference time. In both scenarios, the total power is dominated by the contribution of the logic part, which includes the clock tree as well as the memory-muxing logic (Fig.~\ref{Fig_elec_charac}(b)). Nonetheless, the non-scaling I/O and DRAM contribution increases at lower core supply, consuming more than a third of the total power. This contribution could further worsen for workloads with larger 3D layers, in which case more EMAs per event are necessary.
In contrast, logic related to the datapath of the accelerator's engines represents less than 20\% of the total chip area, which is dominated by SRAMs. This underlines the high storage capacity required by state-of-the-art deep EV-GNN models, as a result from the inherent state-based nature of the graph.

\subsection{Performance Evaluation}
We also evaluate the impact of different parameter configurations on the chip's performance when running the DAGr-GNN workload, at both the accelerator and system levels.

\begin{figure}[!t]
    \centering
    \includegraphics[width=\linewidth]{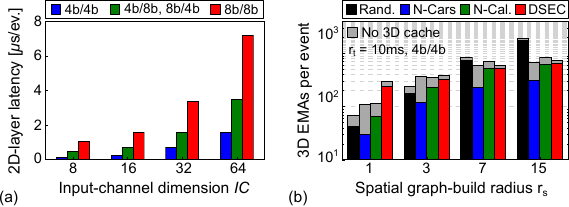}
    \caption{(a) Trade-off between latency per event, X/W operands precision and input channel ($IC$) depth for a single 2D layer ($OC = 64$, 10$^3$ events, 0.95V). (b) Evaluation of 3D EMAs for different spatial radii $r_s$, with and without caching ($IC = 16$ and $r_t = 10$ms).}
    \label{Fig_accelerator_eval}
\end{figure}

At the accelerator level, we investigate the performance of standalone 2D and 3D layers. On the one hand, Fig.~\ref{Fig_accelerator_eval}(a) reports the evolution in execution time of a single 2D layer for different numbers of input-feature channels $IC$ and operand precisions, averaged on 10$^3$ 2D inputs with a varying neighbor sparsity. We demonstrate a quasi-linear scalability of the performance with both parameters, especially at a high $IC$, where the spline convolution's overall execution time is dominated by the MP phase. 
On the other hand, Fig.~\ref{Fig_accelerator_eval}(b) showcases the evolution of EMAs with different graph-building radii $r_s$. These EMAs include the transfer of valid neighbors and their features, closely related to the cache hit rate predictions in Fig.~\ref{Fig_mem_perf}(a), but also account for EMAs related to invalid neighbors, as well as write-backs from the cache. The combination of these effects leads to a maximum effective EMA reduction of 60\% on N-CARS, 33\% on N-Caltech101, and 15\% on DSEC compared to a cacheless baseline. 
Besides, the EMA count increases with $r_s$, with a different rate per dataset. This difference arises from the discrepancy in DVS camera resolution, spatiotemporal locality and average number of neighbors per event. Random inputs have the largest increase in EMAs with $r_s$: their randomness leads to a low average neighbor count, such that the graph-building process rarely stops before reaching the end of the spiral radius, inducing a superlinear increase. In contrast, the higher neighborhood locality of the three edge-vision datasets leads to a more frequent saturation of the maximum neighbor count, early-stopping the graph-building process. The neighborhood density notably increases with the DVS resolution: for DSEC, high EMAs are observed at a small radius due to the feature transfers of several neighbors, but the increase rate lowers because of a rapid saturation of the neighbor count.

\begin{figure}[!t]
    \centering
    \includegraphics[width=\linewidth]{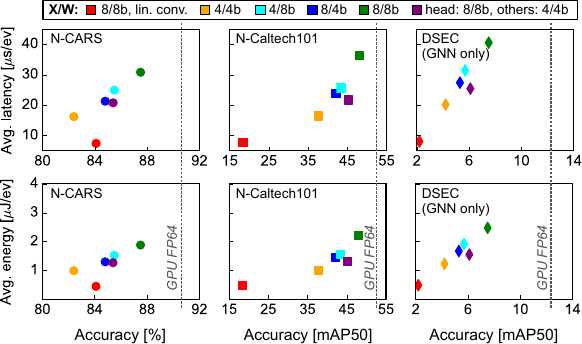}
    \caption{System-level evaluation of DAGr-GNN based on chip measurements, showing the trade-off between classification/detection accuracy, latency and energy per event inference for different convolution types and operand bit precisions (10$^3$ events, 0.95V).}
    \label{Fig_system_eval_trade_off}
\end{figure}

\begin{figure}[!t]
    \centering
    \includegraphics[width=\linewidth]{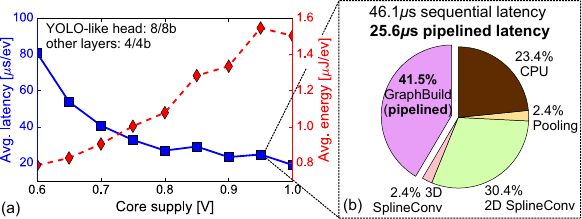}
    \caption{(a) Evolution of the average latency and energy per event inference on DAGr-GNN with supply voltage, at the maximum operating frequency. (b) End-to-end latency breakdown per operation type across DAGr-GNN at 0.95V (10$^3$ events).}
    \label{Fig_system_eval_breakdown}
\end{figure}

At the system level, we evaluate performance on the end-to-end DAGr-GNN workload (except for the three last layers of the YOLO-like head, evaluated off-chip with full accuracy) and average the results over ten sample streams of $10^3$ events each. We first evaluate the trade-off between classification/detection accuracy and the latency (resp. energy) per event inference on different edge-vision datasets. Across all three datasets, spline convolution provides a significant accuracy advantage over linear convolution at iso-precision. Still, it comes at the cost of a 3.8$\times$ latency and energy per event penalty. We can nevertheless play with the trade-off's balance thanks to ETHEREAL's support for mixed 4/8b operands. In particular, we find one of the best trade-offs to be an 8b/8b YOLO-head combined with 4b/4b other layers, respectively achieving a 25.6$\mu$s latency and 1.6$\mu$J per event inference.

We also investigate in Fig.~\ref{Fig_system_eval_breakdown} how system-level performance is impacted by supply voltage, which could serve as a latency-energy turning knob in use cases that have a low or moderate event rate ($\leq 1$Mev/s). This balance is measured in Fig.~\ref{Fig_system_eval_breakdown}(a). Latency improvements closely follow the quasi-linear increase in maximum core-clock frequency with supply voltage (from Fig.~\ref{Fig_elec_charac}(a)). A few nonlinearities still appear at high supply voltage as the division factor of the HS I/F clock has to be increased to satisfy the fixed delays of EMAs. An average breakdown of the end-to-end latency per event inference is detailed in Fig.~\ref{Fig_system_eval_breakdown}(b), at 0.95V and 4b/4b layers: although the graph-building contribution dominates, it can be hidden with pipelining (Section V.C) to bring the latency down to 25.6$\mu$s. 2D spline-convolution layers and CPU transfers dominate the rest of the execution: the latter results from the processing of position pooling directly on the CPU to enable a 32b representation of the accumulated mean position. This overhead could however be solved in a future design by adapting the 2D memory accordingly. 
In terms of energy consumption, Fig.~\ref{Fig_system_eval_trade_off}(a) shows a superlinear trend instead of a quadratic evolution with supply voltage. It results from the significant leakage current in the large chip's SRAMs, which increasingly contributes to the total power as supply voltage gets lowered, reaching down to 0.54$\mu$J/ev at 0.6V.

\subsection{Discussion}
We finally compare the performance of ETHEREAL to that of various prior event-based edge-vision processors that involve an asynchronous processing of the event streams. Table \ref{Tab1} summarizes the key metrics of this comparison, where ETHEREAL shines as the first event-based processor to support high-resolution workloads such as DSEC (640$\times$480).

First, a comparison with a GPU-based implementation of the FP64 DAGr-GNN network showcases the ineffectiveness of the GPU to cope with the $\mu$s-level event stream, mainly owing to the irregularity of the memory transfers. By considering hardware constraints at training time, we limit the accuracy drop between this baseline and our mixed 4b/8b-quantized solution below 5mAP on DSEC. 
Moreover, one should note that these accuracy numbers are obtained for a \textit{directed} graph, explaining the gap in accuracy on DAGr-GNN (GNN only) workload compared to values in \cite{Gehrig2024} at an iso network size. Indeed, these values were obtained for a \textit{bidirectional} graph, leveraging a better spatial information at the cost of a 10$\times$ computational overhead, which makes our low-latency target unreachable. However, \cite{Gehrig2024} demonstrates that the combination of image and event data strongly limits the accuracy drop with a directed GNN, thus combining the fast update of the directed EV-GNN with the structure of the CNN information. This eventual workload is the reason ETHEREAL was optimized for directed graphs.

Compared to both off-the-shelf \cite{Viale2021} and custom \cite{Frenkel2022, Fang2025} SNN processors for vision tasks, ETHEREAL achieves a 20-to-100$\times$ improvement in latency per inference and at least a 2.5$\times$ lower energy. These improvements are obtained while running a deeper network acting on a higher-resolution input, which further underlines their significance.

Compared to prior EV-GNN processors, ETHEREAL achieves comparable accuracy and performance levels on tasks with a low input resolution (e.g., N-CARS and N-Caltech101), while being the only work to provide a scalable architecture to support high-resolution maps.
On the one hand, \cite{Jeziorek2025} surpasses ETHEREAL's detection accuracy on N-Caltech101 by 14mAP, but this improvement stems from a temporal-windowing stage that combines several nodes across space and time into a synchronous graph, as well as from the use of bidirectional edges. As a result, the actual per-event inference latency of this design is orders-of-magnitude higher than ETHEREAL's, and is therefore unable to provide a real-time low-latency response.
On the other hand, \cite{Yang2025} and \cite{Liu2026} both achieve a lower latency per event inference than ETHEREAL, but are specific few-layer architectures dedicated to the toy N-CARS use case. In practice, both architectures suffer from scalability issues: \cite{Yang2025} encodes the space-time information $(x, y, t)$ of past events in on-chip SRAMs, which would amount to hundreds of MBs of on-chip memory for high-resolution tasks. Moreover, their accelerator is limited to sequential neighbor and input-channel processing, and entirely relies on a DRAM storage of graph features. 
While \cite{Liu2026} supports a form of neighbor-parallel event processing and brings interesting message-reuse ideas, they still rely on simple linear kernels and completely discard past events that cannot fit on their $\sim$512kB total on-chip memory, in a FIFO-like manner. While they showcase only a small drop in inference accuracy on N-CARS with this strategy, it likely implies significant accuracy degradation in use cases where different temporal scales come into play, such as DSEC, raising concerns regarding HW scalability.

Altogether, ETHEREAL offers a best-in-class trade-off between flexibility, performance and scalability. Nonetheless, it also leaves several avenues open for future work. First, to leverage higher hit rates, the 3D cache associativity-versus-overhead trade-off should be further studied, together with policies enabling a dynamic allocation of cache resources. This could prove pivotal for workloads with several, many-channel 3D layers, such as motion prediction \cite{dampfhoffer2025graph}, which would otherwise be strongly DRAM-bounded. Second, the MP cores of the spline-convolution engine suffer from under-utilization when neighbor sparsity is high, which could be alleviated through event-level parallelism at the same timestamp. To further improve parallelism, in-memory computing \cite{kneip2023impact} could be considered to combine the weight SRAMs with the MP-core datapath, thereby alleviating the WMEM's bandwidth. Third, the actual combination of the EV-GNN core with a CNN core for DAGr-type workload, or with an event-aggregation core for motion-prediction workloads \cite{dampfhoffer2025graph}, entails open system-level challenges such as heterogeneous core scheduling, data reuse and sharing opportunities, and memory splitting. To that end, framework shells for the integration of heterogeneous accelerators, such as \cite{antonio2025open, antonio2026}, could prove to be a strong baseline to extend our current system. 

\begin{table}[!t]
\centering
\renewcommand{\arraystretch}{1.1}
\setlength{\tabcolsep}{3pt}

\caption{Comparison to the state of the art}
\label{Tab1}

\resizebox{0.5\textwidth}{!}{%

\begin{tabular}{
    >{\raggedright\arraybackslash}p{2.0cm}|  
    >{\centering\arraybackslash}p{1.2cm}|    
    >{\centering\arraybackslash}p{1.0cm}    
    >{\centering\arraybackslash}p{0.9cm}    
    >{\centering\arraybackslash}p{1.0cm}|   
    >{\centering\arraybackslash}p{1.0cm}    
    >{\centering\arraybackslash}p{1.2cm}    
    >{\centering\arraybackslash}p{1.2cm}    
    >{\centering\arraybackslash}p{2.2cm}    
}
\toprule
            & \multicolumn{1}{|c|}{\textbf{GPU}} 
            & \multicolumn{3}{c|}{\textbf{SNN}}
            & \multicolumn{4}{c}{\textbf{EV-GNN}} 
            \\ \midrule
            & Measured & {\cite{Viale2021}\textsuperscript{(iii)}} 
            & {\cite{Frenkel2022}\textsuperscript{(iii)}} 
            & {\cite{Fang2025}\textsuperscript{(iii)}} 
            & {\cite{Yang2025}} 
            & {\cite{Jeziorek2025}}
            & {\cite{Liu2026}} & \textcolor{blue}{\textbf{This Work}\textsuperscript{(iii)}}  \\
\midrule

Year               & 2026 & 2021 & 2022 & 2024 & 2025 & 2025 & 2026 & 2026 \\
Technology         & A100 & 14nm & 28nm & 40nm & FPGA & FPGA & FPGA & 28nm \\
Supply [V]         & 0.9 & 0.8  & 0.5-0.8  & --   & 0.8  & 0.8 & 0.8 & 0.6-1 \\
Freq. [MHz]        & -- & 60 & 13-115 & 50-200 & 200 & -- & 200 & 50-280 \\
Total Memory       & -- & 27MB & 138kB & -- & 780kB & -- & 500kB & 1.25MB \\
Area [mm$^2$]      & -- & 60 & 0.9 & 2.2 & -- & -- & -- & 3.7 \\
X/W Precision      & FP64 & 1/8b & 1/8b & 1-8/8b & 8b & 8b & 8/32b & 4-8/4-8b \\
Max. Img. Res.     & 640$\times$480 & 128$\times$128 & 128$\times$128 & 128$\times$128 & \textcolor{red}{120$\times$100} & 240$\times$180 & \textcolor{red}{120$\times$100} & \textbf{\color{blue}{640$\times$480}} \\
\midrule

\, \newline \textit{Repr. workload} \newline \newline Acc. on HW \newline [\% or mAP50]
&
\textit{N-CARS} \newline 90.7\textsuperscript{(iv)}  
\textit{N-Cal.101} \newline 52.6\textsuperscript{(iv)}  
\textit{DSEC} \newline 12.4\textsuperscript{(iv)} \newline 35.7\textsuperscript{(v)}
& 
\textit{N-CARS} \newline 94.5 \newline \textit{DVS128} \newline 90.5 \newline \, \newline -- \newline \,
&
-- \newline \, \newline \textit{DVS128}\newline 87.3 \newline \, \newline -- \newline \,
&
-- \newline \, \newline \textit{DVS128}\newline 96.1 \newline \, \newline -- \newline \,
&
\textit{N-CARS} \newline 87.8 \newline \, \newline -- \newline \, \newline -- 
&
\textit{N-CARS} \newline 92.5\newline
\textit{N-Cal.101} \newline \textcolor{blue}{62.8} \newline \, \newline -- \newline \,
&
\textit{N-CARS} \newline 88.1\newline
-- \newline \, \newline \, \newline -- \newline \,
&
\textit{N-CARS} \newline 86.7\textsuperscript{(iv),(vi),(vii)}\newline
\textit{N-Cal.101} \newline 48.1\textsuperscript{(iv),(vi),(vii)}\newline
\textit{DSEC} \newline 7.6\textsuperscript{(iv),(vi),(vii)} \newline 30.3\textsuperscript{(v),(vi),(vii)}
\\
\midrule

Peak Thrput.\newline [TOPS/b]\textsuperscript{(i),(ii)}
& -- & 0.04 & 0.04 & 92 & -- & 0.24 & -- & 16 \\

Peak Ene. Eff. \newline [TOPS/W/b]\textsuperscript{(i),(ii)}
& -- & 0.48 & 12 & 2740 & -- & -- & -- & 113 \\

Avg. Latency/inf.
& \textcolor{red}{140ms}
& 900$\mu$s 
& 600$\mu$s 
& \textcolor{red}{2.2ms} 
& 16$\mu$s
& \textcolor{red}{$\geq$ 5.1ms}
& \textcolor{blue}{0.6$\mu$s}
& \textbf{\textcolor{blue}{20-40$\mu$s\textsuperscript{(iv),(vi)}}} \\

Avg. Energy/inf.
& \textcolor{red}{400mJ} & 320$\mu$J & 46$\mu$J & 6.2$\mu$J & -- & -- & -- & \textbf{\textcolor{blue}{1.2-2.5$\mu$J\textsuperscript{(iv),(vi)}}} \\
\bottomrule
\end{tabular}

}


\flushleft

\scriptsize{
(i) Linearly normalized to 1b inputs and weights.\,
(ii) 1 MAC = 2 OPs.\,
(iii) System level (excl.\ DRAM).\,
(iv) DAGr-GNN, GNN only. \,
(v) DAGr-GNN, GNN + CNN (FP64). \,
(vi) QAT: 8b/8b head and 4b/4b other layers (0.95V).
(vii) End-to-end acc.\ simulated with HW model, measured five-layer validity with dataset.
}

\end{table}

\section{Conclusion}
\label{sect:conclusion}

In this work, we proposed ETHEREAL, the first event-driven GNN processor chip for low-latency edge-vision applications that scales to high-resolution DVS cameras.
To leverage both the sparse-irregular memory and dense-regular compute requirements of state-of-the-art EV-GNN workloads, ETHEREAL simultaneously introduces a neighbor-parallel spline-convolution datapath with flexible 4/8b precision per operand, as well as a 1.25MB split 3D/2D memory hierarchy. The latter reduces EMAs by up to 60\% in 3D and avoids them altogether in 2D, resulting in a 57$\times$ latency reduction compared to prior serial-based approaches.
Measurement results on the state-of-the-art DAGr-GNN workload showcase an average end-to-end latency (resp. energy) per event inference ranging between 20.1 and 40.1$\mu$s (resp. 1.2 and 2.5$\mu$J) at 0.95V, increasing with the bit-precision configuration. These results demonstrate the scalability of ETHEREAL, which achieves a tenfold improvement over prior scalable designs at iso task complexity.

\section*{Acknowledgments}
This work was funded in part by the Dutch government and by Prophesee
as an HTSM-TKI project. Circuit fabrication was supported by TSMC via the University Shuttle Program.

\bibliographystyle{IEEEtran}
\bibliography{ref}

@inproceedings{Maqueda2018,
author = {Maqueda, Ana I. and Loquercio, Antonio and Gallego, Guillermo and García, Narciso and Scaramuzza, Davide},
title = {{Event-Based Vision Meets Deep Learning on Steering Prediction for Self-Driving Cars}},
booktitle = {Proceedings of the IEEE Conference on Computer Vision and Pattern Recognition (CVPR)},
pages={5419--5427},
year = {2018}
}

@article{Gupta2021,
  author    = {Gupta, A. and Anpalagan, A. and Guan, L. and Khwaja, A. S.},
  title     = {{Deep learning for object detection and scene perception in self-driving cars: Survey, challenges, and open issues}},
  journal   = {Array (Elsevier)},
  volume    = {10},
  pages     = {1--20},
  year      = {2021},
  doi       = {10.1016/j.array.2021.100057}
}

@article{Elbamby2018,
  author    = {Elbamby, Mohammed S. and Perfecto, Cristina and Bennis, Mehdi and Doppler, Klaus},
  journal   = {IEEE Network}, 
  title     = {{Toward Low-Latency and Ultra-Reliable Virtual Reality}}, 
  year      = {2018},
  volume    = {32},
  number    = {2},
  pages     = {78--84},
  doi       = {10.1109/MNET.2018.1700268}
}

@inproceedings{cai2016unified,
  title={{A unified multi-scale deep convolutional neural network for fast object detection}},
  author={Cai, Zhaowei and Fan, Quanfu and Feris, Rogerio S and Vasconcelos, Nuno},
  booktitle={European conference on computer vision},
  pages={354--370},
  year={2016},
  organization={Springer}
}

@inproceedings{cannici2019asynchronous,
  title={{Asynchronous convolutional networks for object detection in neuromorphic cameras}},
  author={Cannici, Marco and Ciccone, Marco and Romanoni, Andrea and Matteucci, Matteo},
  booktitle={Proceedings of the IEEE/CVF Conference on Computer Vision and Pattern Recognition Workshops},
  year={2019}
}

@inproceedings{messikommer2020event,
  title={{Event-based asynchronous sparse convolutional networks}},
  author={Messikommer, Nico and Gehrig, Daniel and Loquercio, Antonio and Scaramuzza, Davide},
  booktitle={European Conference on Computer Vision},
  pages={415--431},
  year={2020},
  organization={Springer}
}

@inproceedings{sironi2018hats,
  title={{HATS: Histograms of averaged time surfaces for robust event-based object classification}},
  author={Sironi, Amos and Brambilla, Manuele and Bourdis, Nicolas and Lagorce, Xavier and Benosman, Ryad},
  booktitle={Proceedings of the IEEE conference on computer vision and pattern recognition},
  pages={1731--1740},
  year={2018}
}

@inproceedings{gehrig2023recurrent,
  title={{Recurrent vision transformers for object detection with event cameras}},
  author={Gehrig, Mathias and Scaramuzza, Davide},
  booktitle={Proceedings of the IEEE/CVF conference on computer vision and pattern recognition},
  pages={13884--13893},
  year={2023}
}

@article{chen2024survey,
  title={{A survey on graph neural networks and graph transformers in computer vision: A task-oriented perspective}},
  author={Chen, Chaoqi and Wu, Yushuang and Dai, Qiyuan and Zhou, Hong-Yu and Xu, Mutian and Yang, Sibei and Han, Xiaoguang and Yu, Yizhou},
  journal={IEEE Transactions on Pattern Analysis and Machine Intelligence},
  volume={46},
  number={12},
  pages={10297--10318},
  year={2024},
  publisher={IEEE}
}

@inproceedings{markidis2018nvidia,
  title={{Nvidia tensor core programmability, performance \& precision}},
  author={Markidis, Stefano and Der Chien, Steven Wei and Laure, Erwin and Peng, Ivy Bo and Vetter, Jeffrey S},
  booktitle={2018 IEEE international parallel and distributed processing symposium workshops (IPDPSW)},
  pages={522--531},
  year={2018},
  organization={IEEE}
}

@article{chen2019eyeriss,
  title={{Eyeriss v2: A flexible accelerator for emerging deep neural networks on mobile devices}},
  author={Chen, Yu-Hsin and Yang, Tien-Ju and Emer, Joel and Sze, Vivienne},
  journal={IEEE Journal on Emerging and Selected Topics in Circuits and Systems},
  volume={9},
  number={2},
  pages={292--308},
  year={2019},
  publisher={IEEE}
}

@article{kneip2023impact,
  title={{IMPACT: A 1-to-4b 813-TOPS/W 22-nm FD-SOI compute-in-memory CNN accelerator featuring a 4.2-POPS/W 146-TOPS/mm 2 CIM-SRAM with multi-bit analog batch-normalization}},
  author={Kneip, Adrian and Lefebvre, Martin and Verecken, Julien and Bol, David},
  journal={IEEE Journal of Solid-State Circuits},
  volume={58},
  number={7},
  pages={1871--1884},
  year={2023},
  publisher={IEEE}
}

@inproceedings{dumoulin2024enabling,
  title={{Enabling Efficient Hardware Acceleration of Hybrid Vision Transformer (ViT) Networks at the Edge}},
  author={Dumoulin, Joren and Houshmand, Pouya and Jain, Vikram and Verhelst, Marian},
  booktitle={2024 IEEE International Symposium on Circuits and Systems (ISCAS)},
  pages={1--5},
  year={2024},
  organization={IEEE}
}

@inproceedings{dong202528nm,
  title={{A 28nm 0.22 $\mu$j/token memory-compute-intensity-aware cnn-transformer accelerator with hybrid-attention-based layer-fusion and cascaded pruning for semantic-segmentation}},
  author={Dong, Pingcheng and others},
  booktitle={2025 IEEE International Solid-State Circuits Conference (ISSCC)},
  volume={68},
  pages={01--03},
  year={2025},
  organization={IEEE}
}

@article{gehrig2021dsec,
  title={{Dsec: A stereo event camera dataset for driving scenarios}},
  author={Gehrig, Mathias and Aarents, Willem and Gehrig, Daniel and Scaramuzza, Davide},
  journal={IEEE Robotics and Automation Letters},
  volume={6},
  number={3},
  pages={4947--4954},
  year={2021},
  publisher={IEEE}
}

@inproceedings{antonio2025open,
  title={{An open-source hw-sw co-development framework enabling efficient multi-accelerator systems}},
  author={Antonio, Ryan Albert and Dumoulin, Joren and Yi, Xiaoling and Van Delm, Josse and Deng, Yunhao and Paim, Guilherme and Verhelst, Marian},
  booktitle={2025 IEEE/ACM International Symposium on Low Power Electronics and Design (ISLPED)},
  pages={1--7},
  year={2025},
  organization={IEEE}
}

@inproceedings{schaefer2022aegnn,
  title={{Aegnn: Asynchronous event-based graph neural networks}},
  author={Schaefer, Simon and Gehrig, Daniel and Scaramuzza, Davide},
  booktitle={Proceedings of the IEEE/CVF conference on computer vision and pattern recognition},
  pages={12371--12381},
  year={2022}
}

@inproceedings{dampfhoffer2025graph,
  title={{Graph Neural Network Combining Event Stream and Periodic Aggregation for Low-Latency Event-based Vision}},
  author={Dampfhoffer, Manon and Mesquida, Thomas and Joubert, Damien and Dalgaty, Thomas and Vivet, Pascal and Posch, Christoph},
  booktitle={Proceedings of the IEEE/CVF Conference on Computer Vision and Pattern Recognition},
  pages={6909--6918},
  year={2025}
}

@inproceedings{Verma2024,
  author    = {Verma, Aayush A. and Chakravarthi, Bharatesh and Vaghela, Arpitsinh and Wei, Hua and Yang, Yezhou},
  title     = {{ETraM: Event-Based Traffic Monitoring Dataset}},
  booktitle = {Proceedings - 2024 IEEE/CVF Conference on Computer Vision and Pattern Recognition, CVPR 2024},
  pages     = {22637--22646},
  year      = {2024},
  doi       = {10.1109/CVPR52733.2024.02136}
}

@article{Lichtsteiner2008,
  author    = {Lichtsteiner, Patrick and Posch, Christoph and Delbruck, Tobi},
  journal   = {IEEE Journal of Solid-State Circuits}, 
  title     = {{A 128$\times$128 120 dB 15 $\mu$s Latency Asynchronous Temporal Contrast Vision Sensor}}, 
  year      = {2008},
  volume    = {43},
  number    = {2},
  pages     = {566--576},
  doi       = {10.1109/JSSC.2007.914337}
}

@article{sauter2025croc,
  title={{Croc: An end-to-end open-source extensible risc-v mcu platform to democratize silicon}},
  author={Sauter, Phillippe and Benz, Thomas and Scheffler, Paul and Pochert, Hannah and W{\"u}thrich, Luisa and Povi{\v{s}}er, Martin and Muheim, Beat and G{\"u}rkaynak, Frank K and Benini, Luca},
  journal={arXiv preprint arXiv:2502.05090},
  year={2025}
}

@inproceedings{Viale2021,
  author    = {Viale, Alberto and Marchisio, Alberto and Martina, Maurizio and Masera, Guido and Shafique, Muhammad},
  booktitle = {2021 International Joint Conference on Neural Networks (IJCNN)}, 
  title     = {{CarSNN: An Efficient Spiking Neural Network for Event-Based Autonomous Cars on the Loihi Neuromorphic Research Processor}}, 
  year      = {2021},
  volume    = {},
  number    = {},
  pages     = {1--10},
  doi       = {10.1109/IJCNN52387.2021.9533738}
}

@inproceedings{Frenkel2022,
  author    = {Frenkel, Charlotte and Indiveri, Giacomo},
  booktitle = {2022 IEEE International Solid-State Circuits Conference (ISSCC)}, 
  title     = {{ReckOn: A 28nm Sub-mm2 Task-Agnostic Spiking Recurrent Neural Network Processor Enabling On-Chip Learning over Second-Long Timescales}}, 
  year      = {2022},
  volume    = {65},
  number    = {},
  pages     = {1--3},
  doi       = {10.1109/ISSCC42614.2022.9731734}
}

@article{Fang2025,
  author    = {Fang, Chaoming and Shen, Ziyang and Wang, Zongsheng and Wang, Chuanqing and Zhao, Shiqi and Tian, Fengshi and Yang, Jie and Sawan, Mohamad},
  journal   = {IEEE Journal of Solid-State Circuits}, 
  title     = {{An Energy-Efficient Unstructured Sparsity-Aware Deep SNN Accelerator With 3-D Computation Array}}, 
  year      = {2025},
  volume    = {60},
  number    = {3},
  pages     = {977--989},
  doi       = {10.1109/JSSC.2024.3507095}
}

@article{Yang2025,
  author    = {Yang, Yufeng and Kneip, Adrian and Frenkel, Charlotte},
  journal   = {IEEE Transactions on Circuits and Systems for Artificial Intelligence}, 
  title     = {{EvGNN: An Event-Driven Graph Neural Network Accelerator for Edge Vision}}, 
  year      = {2025},
  volume    = {2},
  number    = {1},
  pages     = {37--50},
  doi       = {10.1109/TCASAI.2024.3520905}
}

@article{Gehrig2024,
  author    = {Gehrig, Daniel and Scaramuzza, Davide},
  title     = {{Low-latency automotive vision with event cameras}},
  journal   = {Nature},
  volume    = {629},
  pages     = {1034--1040},
  year      = {2024},
  doi       = {10.1038/s41586-024-07409-w}
}

@article{Jeziorek2025,
  author    = {Jeziorek, Kamil and Wzorek, Piotr and Blachut, Krzysztof and Pinna, Andrea and Kryjak, Tomasz},
  title     = {{Embedded Graph Convolutional Networks for Real-Time Event Data Processing on SoC FPGAs}},
  journal   = {arXiv preprint arXiv:2406.07318},
  pages     = {1--16},
  year      = {2025},
  doi       = {10.48550/arXiv.2406.07318}
}

@inproceedings{Matthias2018,
  author    = {Fey, Matthias and Lenssen, Jan Eric and Weichert, Frank and Müller, Heinrich},
  title     = {{SplineCNN: Fast Geometric Deep Learning With Continuous B-Spline Kernels}},
  booktitle = {Proceedings of the IEEE Conference on Computer Vision and Pattern Recognition (CVPR)},
  month     = {June},
  year      = {2018},
  doi       = {10.48550/arXiv.1711.08920}
}

@article{Liu2026,
  author    = {Liu, Tianhang and Zhang, Shen and Yan, Guangyao and Wang, Runhua and Li, Rui and Meng, Shijie and Sun, Hao and Ha, Yajun},
  journal   = {IEEE Transactions on Circuits and Systems I: Regular Papers}, 
  title     = {{Event-Driven Asynchronous Graph Neural Network FPGA Accelerator for Real-Time Edge Vision}}, 
  year      = {2026},
  volume    = {},
  number    = {},
  pages     = {1--14},
  doi       ={10.1109/TCSI.2026.3679753}
}

@inproceedings{antonio2026,
  author    = {Antonio, Ryan Albert and Dumoulin, Joren and Yi, Xiaoling and Feng, Jun and Deng, Yunhao and Kong, Fanchen and Kneip, Adrian and Paim, Guilherme and Dehaene, Wim and Verhelst, Marian},
  booktitle = {2026 IEEE European Solid-State Electronics Research Conference (ESSERC) (accepted)}, 
  title     = {{Hemaia: a Hetereogeneous Multi-Accelerator System-on-Chip}}, 
  year      = {2026},
  volume    = {},
  number    = {},
  pages     = {1--4},
  doi       = {}
}

@ARTICLE{ottati2023_spiking,
  author={Ottati, Fabrizio and Gao, Chang and Chen, Qinyu and Brignone, Giovanni and Casu, Mario R. and Eshraghian, Jason K. and Lavagno, Luciano},
  journal={IEEE Journal on Emerging and Selected Topics in Circuits and Systems}, 
  title={{To Spike or Not to Spike: A Digital Hardware Perspective on Deep Learning Acceleration}}, 
  year={2023},
  volume={13},
  number={4},
  pages={1015--1025},
  doi={10.1109/JETCAS.2023.3330432}
}

@inproceedings{cacti_2008,
  author={Thoziyoor, Shyamkumar and Ahn, Jung Ho and Monchiero, Matteo and Brockman, Jay B. and Jouppi, Norman P.},
  booktitle={2008 International Symposium on Computer Architecture}, 
  title={{A Comprehensive Memory Modeling Tool and Its Application to the Design and Analysis of Future Memory Hierarchies}}, 
  year={2008},
  volume={},
  number={},
  pages={51-62},
  doi={10.1109/ISCA.2008.16}
}

@article{camera_2026,
  author={Shi, Xiaopei and Ji, Ruoyu and Nie, Kaiming and Gao, Zhiyuan and Gao, Jing and Xin, Yu and Xu, Jiangtao},
  journal={IEEE Transactions on Circuits and Systems I: Regular Papers}, 
  title={{A 40k fps 99.5 dB Wide Dynamic Range PWM Image Sensor Featuring a Low-Power Pixel Array}}, 
  year={2026},
  volume={},
  number={},
  pages={1-12},
  doi={10.1109/TCSI.2026.3695072}
}

\end{document}